\pdfoutput=1
\documentclass[aps,prl,reprint,superscriptaddress,nofootinbib]{revtex4-2}
\usepackage{color}
\usepackage{amsmath}
\usepackage{amssymb}
\usepackage{graphicx}
\usepackage{booktabs}
\usepackage{bm}
\graphicspath{{figures/}}

\usepackage{hyperref}
\definecolor{linkcolor}{rgb}{0.0,0.3,0.5}
\hypersetup{colorlinks=true,linkcolor=linkcolor,citecolor=linkcolor,
            filecolor=linkcolor,urlcolor=linkcolor}

\makeatletter
\newcommand\mysec[1]{\vspace{2mm}\noindent\def\@currentlabel{#1}\emph{#1}---}
\makeatother

\newcommand\dT{\delta_T}
\newcommand\Vkd{V^{\rm KD}}

\begin{document}

\preprint{LA-UR-26-28459}
\title{Rapid Uncertainty Quantification on a Latent Field using Fisher Information}
\author{ Karl Daningburg }
\email{kd1956@rit.edu}
\affiliation{ School of Mathematics and Statistics, Rochester Institute of Technology, Rochester, New York, 14623}
\affiliation{ Theoretical Division, Los Alamos National Laboratory, Los Alamos, New Mexico, 87545 }
\author{ A.~E.~Lovell }
\email{lovell@lanl.gov}
\affiliation{ Theoretical Division, Los Alamos National Laboratory, Los Alamos, New Mexico, 87545 }
\author{ Arvind T.~Mohan }
\email{arvindm@lanl.gov}
\affiliation{ Computational Physics and Methods Group, Los Alamos National Laboratory, Los Alamos, New Mexico, 87545 }
\author{ R.~O'Shaughnessy}
\affiliation{ School of Mathematics and Statistics, Rochester Institute of Technology, Rochester, New York, 14623}

\begin{abstract}
Many inverse problems in physics infer an unobserved field from
measurements connected to it through a governing equation.
We combine Fisher information with a differentiable solver to
quantify uncertainty in both the inferred field and its predicted
observables, using a local Gaussian approximation without sampling
the full parameter posterior.
Applied to nuclear optical potentials, the method distinguishes
features constrained by scattering measurements from those whose
uncertainty remains set by the prior.
Our flexible model improves agreement with held-out cross sections
and achieves nearly nominal coverage with narrower predictive
intervals than an established Bayesian optical model.
\end{abstract}
\maketitle

\mysec{Introduction}
Many inverse problems in physics seek to infer an unobserved field
from measurements of observables connected to that field through
a governing equation~\cite{backus1968,stuart2010inverse}. In nuclear reaction theory, 
the optical potential is such a field:
an effective interaction between a projectile and a target nucleus
inferred from scattering measurements. In the optical model, the elastic and inelastic interaction is captured by solving the
Schr\"odinger equation with a complex-valued potential. Optical models are widely used in reaction modelling~\cite{hebborn2023rib},
both directly in the analysis of experimental data and as inputs to
downstream models ranging from nuclear
fission~\cite{beyer2025fissionuq} to the $r$-process, the rapid capture
of neutrons that builds roughly half the elements heavier than
iron~\cite{mumpower2016rprocess}. Since their first use in 1954 \cite{1954PhRv...96..448F},
optical models have relied on assumed functional forms with a limited number
of tunable parameters. These forms serve to constrain the models and make uncertainty
quantification (UQ) via posterior sampling tractable; however,
they restrict the model's ability to fit experimental data.
A further challenge is to quantify the uncertainty of predictions
for nuclei and energies beyond those constrained by
measurements~\cite{hebborn2023rib,beyer2026eastlansing}.

The optical model for neutron scattering captures the interaction between a nucleus and a neutron as a complex
potential built from several components $c$. Central components take a
Woods--Saxon form \cite{1954PhRv...95..577W}, $f(r,R,a) = [1+e^{(r-R)/a}]^{-1}$, with $R = r_0A^{1/3}$ the
nuclear radius and $a$ the surface diffuseness. Surface-peaked components use its
radial derivative, $-4a\,\mathrm{d}f/\mathrm{d}r$, which concentrates strength
at the nuclear surface.
Spin--orbit components take the Thomas form, $r^{-1}\,\mathrm{d}f/\mathrm{d}r$,
also surface-peaked, which couples the neutron's spin to its orbital motion. The Koning--Delaroche (KD) optical model, widely used since 
its formulation in 2003, builds on this form by fitting 46 parameters describing depths, 
radii, and diffuseness as functions of nuclear mass $A$, charge number $Z$, and incident neutron energy $E$~\cite{koning2003}.
Pruitt \emph{et al.} modernized KD (KDUQ) by calibrating the same functional form in a
Bayesian framework, yielding a posterior over its 46 parameters from which the
uncertainty on any predicted observable can be calculated from samples of the parameter space~\cite{pruitt2023kduq}.
The uncertainty predicted by KDUQ and similar models~\cite{beyer2026eastlansing} 
is conditional on the functional forms of the model. Furthermore, the uncertainties 
are not responsive to distance from training data. 

Building on Ref.~\cite{daningburg2026thesis}, we combine a flexible
representation of deviations from an established optical potential
with Fisher-based uncertainty propagation through a differentiable
scattering solver. This allows us to determine which changes in the
radial potential are constrained by measured cross sections and
which remain controlled by the prior. We use the autodifferentiable framework of Ref.~\cite{daningburg2026thesis} 
to propagate uncertainties upstream from data to parameter space, as well 
as downstream from parameters to the potential, and finally to observables predicted by the model. Our method 
applies wherever a field is inferred through a differentiable forward model, avoids the poor scaling of Monte Carlo methods, and yields uncertainties 
that grow where data is sparse.

\mysec{The data and objective}
\label{sec:data}
We fit neutron elastic differential cross sections as listed by Pruitt \emph{et al.} for KDUQ and extracted from EXFOR~\cite{pruitt2023kduq}.
We divide this ``KDUQ Corpus" into a Training split and Validation split. The Training split contains 8298 measured points spanning 272
distinct combinations of target and reaction energy; we will refer to these distinct combinations of nuclear charge, nuclear weight, and 
reaction energy $(Z,A,E)$ as \textit{experiments}. A further 68 experiments,
2227 points, fall into the Validation split and are used to examine performance of our model on unseen data. As an out-of-distribution test
we use the separate ``Test Corpus" of the same work, which Pruitt \emph{et al.}
describe as representing a somewhat different underlying distribution in target nucleus and energy. To avoid compound 
elastic scattering effects and thereby 
defray the need for a Hauser-Feschbach solver, we filter the datasets on a 
per-nucleus basis according to a method described in Ref.~\cite{daningburg2026thesis}.

Each measurement carries a reported uncertainty $\sigma_{{\rm rep},i}$. However, Pruitt \emph{et al.} and
others have noted that these uncertainties are too small to explain the scatter between 
nearby data. We
therefore add a common term $\dT$, following the form of
Ref.~\cite{pruitt2023kduq}, giving
\begin{equation}
  \sigma_i^2 = \sigma_{{\rm rep},i}^2 + \dT^2,
  \qquad
  \sigma_{{\rm pred},i}^2 = \sigma_i^2 + \sigma_{{\rm model},i}^2 .
  \label{eq:errormodel}
\end{equation}
The variance entering the fit is $\sigma_i^2$; the variance a prediction is scored against is $\sigma_{{\rm pred},i}^2$ and includes the model's own uncertainty
$\sigma_{\rm model}$. We calculate
$\dT = 0.137$ from 22 cases in which two groups measured nearly the same
target at nearly the same energy; End Matter has more details 
on this calculation. Pruitt \emph{et al.} fit $\dT$ along with the
potential parameters; their posterior for neutron elastic differential cross sections is
$\dT = 0.20$, and we score KDUQ at that value.
Writing $r_i = \ln y_i - \ln \hat y_i$ for the residual of the
$i$th measurement of the logarithm of the cross section, and taking the residuals to be
Gaussian with the variances of Eq.~\eqref{eq:errormodel}, the log-likelihood is $-\chi^2/2$
up to an additive constant, with $\chi^2 \equiv \sum_i (r_i/\sigma_i)^2$. The fit maximizes
this likelihood together with the prior introduced in Methods.

\begin{figure}
  \includegraphics{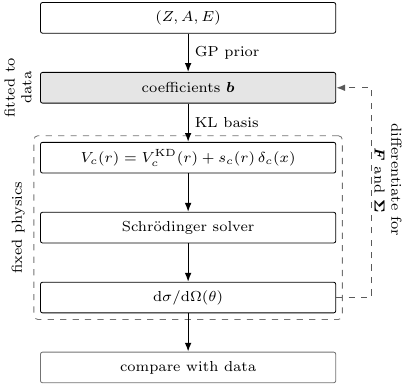}%
  \caption{\label{fig:pipeline}The forward pass. The coefficients $\bm b$ fix the deviation of the
  potential from KD through the Karhunen--Lo\`eve basis of Eq.~\eqref{eq:model}. The potential enters a Schr\"odinger 
  solver that returns the cross section to be compared with the data. Differentiating this chain gives the Fisher information of Eq.~\eqref{eq:fisher} and the posterior covariance of
  Eq.~\eqref{eq:posteriorcov}.}
\end{figure}

\mysec{Methods}
In Ref.~\cite{daningburg2026thesis} we demonstrate that the nuclear optical potential could be expressed 
as a neural network that outputs potential values as a function of radius. 
Here, we instead represent the potential as a deviation
from KD drawn from a Gaussian process (GP), following Ref.~\cite{melendez2019buqeye}:

\begin{equation}
  \delta_c(x) \;\equiv\; \frac{V_c(r) - \Vkd_c(r)}{s_c(r)}
    \;\sim\; \mathcal{GP}\bigl(0,\,k\bigr)
  \label{eq:prior}
\end{equation}
where $x \equiv r/A^{1/3}$ is scaled to the size of the nucleus,
 $V_c(r)$ is the $c$th of the $N_c = 4$ components of the potential (real and imaginary central, the latter
including KD's surface absorption; real and imaginary spin--orbit), $\Vkd_c$ is the equivalent prediction from
KD, and $\mathcal{GP}\bigl(0,\,k\bigr)$ is a Gaussian process prior with mean 0 and Mat\'ern kernel $k$, whose
correlation length in $x$ is set by the nuclear surface diffuseness. 
$s_c$ is a scaling factor which ensures the deviations from KD are appropriate for the 
different optical potential components; more details are in End Matter.
We expand $\delta_c$ in the Karhunen--Lo\`eve (KL) basis of that prior \cite{Levy2008};
the KL modes are uncorrelated, so truncating the expansion discards a known
fraction of the prior variance.
Truncating that expansion at $M$ modes,
\begin{equation}
  \delta_c(x) \;=\; \sum_{j=1}^{M} \sqrt{\lambda_j}\; b_{cj}\, \varphi_j(x),
  \label{eq:model}
\end{equation}
where $\varphi_j$ and $\lambda_j$ are the eigenfunctions and eigenvalues of $k$.
Factoring $\sqrt{\lambda_j}$ out of the coefficient leaves
$b_{cj}$ independent and of unit variance under the prior.
The coefficients $b$ --- $M$ modes in each of the $N_c$ components --- fix the
potential at a given nucleus and energy.
Figure~\ref{fig:pipeline} outlines the algorithm:  mirroring previous work on machine learning for physical
systems~\cite{mohan2023hard}, we adopt a real physics layer (here, the Schr\"odinger solver) to connect our model to the data.

We fit the coefficients $b^{(n)}_{cj}$ of Eq.~\eqref{eq:model}
at each of the $N_e = 272$ measured target--energy combinations
$(A_n,Z_n,E_n)$, giving $N_c M N_e$ parameters. A second GP
specifies how these coefficients vary between nuclei and energies.
Its Mat\'ern kernel favors similar values at nearby points in
standardized $(A,Z,E)$, with the range of this similarity set by
the correlation length.

Measurements therefore constrain deviations from KD both locally and at nearby nuclei and energies. Beyond the correlation length,
their influence diminishes and the posterior approaches the
prior: the mean correction returns to zero, recovering KD,
while its uncertainty returns to the prior width.

\begin{figure*}
  \includegraphics[width=\textwidth]{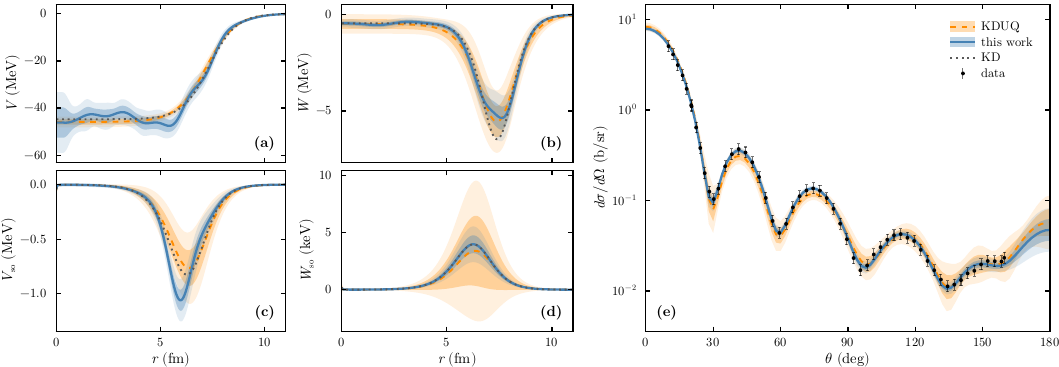}%
  \caption{\label{fig:band}Potential and
  elastic cross section for neutrons on $^{209}$Bi at 10~MeV, from KDUQ (orange) and from
  this work (blue). (a) real and (b) imaginary central potential; (c) real and (d) imaginary
  spin--orbit potential; (e) elastic differential cross section. Bands show the model uncertainty $\sigma_{\rm model}$ (dark
  $\pm1\sigma$, light $\pm2\sigma$). Error bars on the data combine the reported uncertainty with our measured
  $\dT = 0.137$ (Eq.~\eqref{eq:errormodel}).
  }
\end{figure*}
\begin{table*}
\caption{\label{tab:fit}Fit quality against each reference model.
$\xi = \chi^2_{\rm ours}/\chi^2_{\rm ref}$ is the $\chi^2$ error ratio for a single
experiment, and $\bar\xi = \sum\chi^2_{\rm ours}/\sum\chi^2_{\rm ref}$ is
its data-weighted counterpart over a corpus. ``improved exp'' is the percent of experiments with $\xi < 1$; medians are
over experiments, with the interquartile range in brackets. Validation data is
in-sample for KDUQ; only the Test Corpus is out of sample for every model.}
\begin{ruledtabular}
\begin{tabular}{lcccccc}
 &  \multicolumn{3}{c}{vs.\ KDUQ} & \multicolumn{3}{c}{vs.\ KD} \\
corpus  & improved exp & $\bar\xi$ & median $\xi$ & improved exp & $\bar\xi$ & median $\xi$ \\
\hline
Training{} \emph{(in sample)} & 100\% & 0.14 & 0.17 [0.09--0.31] & 99\% & 0.13 & 0.18 [0.08--0.32] \\ 
\hline
Validation  & 81\% & 0.40 & 0.39 [0.16--0.77] & 84\% & 0.35 & 0.35 [0.13--0.75] \\ 
\hline
Test Corpus  & 72\% & 1.00 & 0.82 [0.60--1.03] & 73\% & 0.97 & 0.85 [0.66--1.01] \\ 
\end{tabular}
\end{ruledtabular}
\end{table*}

\mysec{Uncertainty from the measured cross sections}
Differential cross sections constrain the potential through their
sensitivity to its parameters. A change in the potential that produces a large
change in the cross section relative to the measurement uncertainty is strongly
constrained; a change with little effect on the cross section is weakly
constrained. For independent Gaussian errors in the logarithm of the cross
section, this sensitivity is described by the Fisher information,
\begin{equation}
  F_{ab} = \sum_i \frac{1}{\sigma_i^2}
    \frac{\partial \ln \hat y_i}{\partial p_a}
    \frac{\partial \ln \hat y_i}{\partial p_b},
  \label{eq:fisher}
\end{equation}
where $p_a$ and $p_b$ are model parameters, $\hat y_i$ is the predicted
differential cross section, and the sum runs over all measured angles in all
experiments. The uncertainties $\sigma_i$ are those defined 
in Eq.~\eqref{eq:errormodel}. We evaluate the derivatives at the fitted
parameters by automatic differentiation through the scattering solver.

Combining the Fisher information with the prior precision
$\bm\Lambda$ gives, in the local Gaussian approximation,
the posterior covariance
\begin{equation}
  \bm\Sigma \simeq (\bm F + \bm\Lambda)^{-1}.
  \label{eq:posteriorcov}
\end{equation}
The prior restricts potential changes that the cross sections alone cannot
determine. Thus a weakly constrained direction retains uncertainty set by
the prior, while a direction to which the measurements are sensitive can be
determined more precisely. 

For any predicted quantity $q$, a small parameter displacement gives
$\delta q \simeq \sum_a (\partial q/\partial p_a)\,\delta p_a$.
Its marginal distribution is therefore approximately Gaussian, with variance
\begin{equation}
  \sigma^2(q) \simeq \sum_{a,b}
    \frac{\partial q}{\partial p_a}\,
    \Sigma_{ab}\,
    \frac{\partial q}{\partial p_b}.
  \label{eq:band}
\end{equation}
Equation~\eqref{eq:band} gives the potential uncertainty for $q=V_c(r)$, and
the model uncertainty of Eq.~\eqref{eq:errormodel} for
$q=\ln(\mathrm{d}\sigma/\mathrm{d}\Omega)$. Without the prior term $\bm\Lambda$, Eqs.~\eqref{eq:posteriorcov} and 
~\eqref{eq:band} are the standard $\chi^2$-covariance propagation used in nuclear theory 
\cite{dobaczewski2014errors} and for optical-model parameters \cite{lovell2017omp, king2019bayesfreq}. 

 We evaluate these expressions in the dimension of the data
rather than of the parameters, using the Woodbury identity, so the full Fisher matrix is never
formed or inverted. The data reduce the prior
covariance only along directions to which the cross sections are sensitive; elsewhere the
uncertainty remains that of the prior~\cite{cui2014lis,spantini2015}. End Matter gives the construction
and tests of the linearization.

\begin{figure}
  \includegraphics[width=\columnwidth]{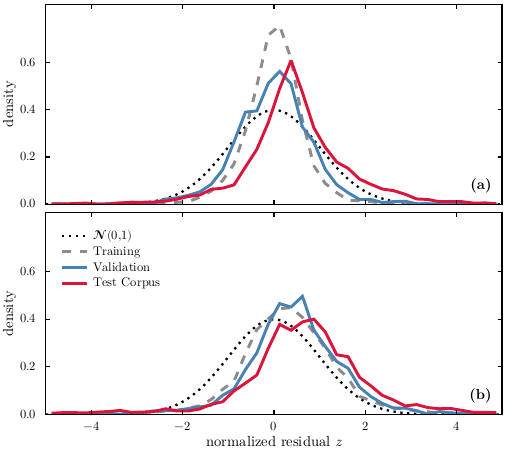}%
  \caption{\label{fig:calib} Normalized residuals of (a) our model and (b) KDUQ on the
  Training, Validation, and Test Corpus data. KDUQ is its published posterior evaluated
  through our solver at their $\dT = 0.2$; Validation data are in-sample for KDUQ. A perfectly
  calibrated model would have a standard normal curve.}
\end{figure}
\begin{table*}
\caption{\label{tab:calib}Calibration of our model{} and KDUQ on identical
data splits. The two $\rho$ columns are our uncertainty in units of
KDUQ's: $\rho = \sigma_\text{ours} / \sigma_\text{KDUQ}$; both are taken
pointwise at each measured angle and the median is taken, with the interquartile
range given for $\rho_{\rm pred}$. KDUQ is scored at its own $\dT = 0.2$, our model at
$\dT = 0.137$. Validation data are withheld from our fit but
lie inside the KDUQ Corpus; only the Test Corpus is out of sample for every
model.}
\begin{ruledtabular}
\begin{tabular}{llccccc}
corpus & model & cov$_{68}$ & cov$_{95}$ & $\rho_{\rm model}$ & $\rho_{\rm pred}$ & rms $z$ \\
\hline
Training & GP & 89.6\% & 98.7\% & $0.47$ & $0.68$ [0.64--0.70] & 0.647 \\ 
\emph{(in sample)}& KDUQ & 69.7\% & 93.0\% & $1$     & $1$ & 1.091 \\
\hline
Validation & GP & 80.4\% & 96.0\% & $0.58$ & $0.70$ [0.67--0.76] & 0.980 \\ 
 & KDUQ & 67.3\% & 92.1\% & $1$     & $1$ & 1.136 \\
\hline
Test Corpus & GP & 67.5\% & 89.3\% & $2.46$ & $1.34$ [0.70--2.24] & 1.214 \\ 
 & KDUQ & 54.7\% & 82.6\% & $1$     & $1$ & 1.696 \\
\end{tabular}
\end{ruledtabular}
\end{table*}

\mysec{Better fits with better calibration}
Figure~\ref{fig:band} shows the potential and the cross section predicted for
$^{209}$Bi at 10~MeV, compared to KDUQ. The experimental data~\cite{das1990bi209} falls in the Validation 
split, which we withheld from training but was used in training KDUQ. Despite this, we predict the cross section both more accurately and more precisely.
Over its 56 measured angles our model gives $\chi^2/N =$
0.26, against
0.80 for KDUQ, and the median $\sigma_{\rm model}$ on the predicted
cross section is 6\% against
KDUQ's 16\%.
Note also the widening of the potential uncertainty, $\sigma(V_c)$ of
Eq.~\eqref{eq:band}, inside the nucleus in panel~(a), a physically sensible result:
interactions within the nucleus at this reaction energy have little impact on the final cross section, so 
our model reflects the uncertainty in the potential in that regime.

Table~\ref{tab:fit} compares $\chi^2$ error among our model, KDUQ, and KD. 
On the Test Corpus --- the only dataset withheld from training for every model --- our
median $\chi^2$ is $0.82$ of KDUQ's, better on
72\% of experiments. On the Validation data, which lie within the KDUQ Corpus, the margin is much larger:
$0.39$ across 81\% of experiments. We emphasize that the 
out-of-distribution Test Corpus is particularly challenging for all models, coming from regions of the nuclear chart
largely unexplored by the KDUQ Corpus. 

Table~\ref{tab:calib} scores our model and KDUQ{} over the three corpora. Two widths are included: the model's own
contribution $\sigma_{\rm model}$ (Eq.~\eqref{eq:band}) and
the full predictive width $\sigma_{\rm pred}$ of
Eq.~\eqref{eq:errormodel}, against which the residuals are scored. We report each as a ratio to KDUQ's,
$\rho_{\rm model} = \sigma_{\rm model}/\sigma_{\rm model}^{\rm KDUQ}$ and
likewise $\rho_{\rm pred}$, so that $\rho < 1$ means our interval is the
narrower one.On data withheld from our fit we have better coverage than KDUQ, with narrower predictive intervals:
our model covers 96.0\% of the nominal 95\% interval
against their 92.1\%, with predictive intervals 30\% narrower.%
. The width columns indicate 
that on these corpora we obtain better coverage through better fits, not wider predictive intervals.

We summarize calibration 
with the normalized residuals $z_i = r_i / \sigma_\text{pred}$, which yields a distribution 
that ideally follows a standard normal distribution, with rms $z=1$. Values of rms $z>1$
indicate uncertainties which are too low for the model's error, whereas rms $z<1$ indicates 
a model which has conservatively wide error bounds. Comparisons with KDUQ contextualize our bounds: KDUQ's residuals are
offset toward positive $z$ on every split, with mean $z$ of 0.34 on
Training and 0.47 on Validation against
0.01 and 0.03
for our model, so KDUQ's rms $z$ exceeds 1 even on the data KDUQ was fit to. KDUQ also degrades
more than our model on the Test Corpus (rms $z=1.696$ versus 1.214).
Our model covers 89.3\% against their 82.6\%, with
$\sigma_{\rm pred}$ $1.34\times$ wider. 
On Validation data our $\sigma_{\rm model}$ is
$0.58$ of KDUQ's, and on the Test Corpus
$2.46$ of KDUQ's: our
$\sigma_{\rm model}$ grows by
$5.22\times$ between the two corpora, while
KDUQ's only grows by $1.23\times$.

Figure~\ref{fig:calib} provides a more resolved look at the normalized residuals.
In panel (a) we evaluate our model's calibration on all three data splits. On data from within the KDUQ Corpus -- Training and Validation -- we show very similar curves which 
are nearly standard normal, but have a higher peak. This indicates that some predictive intervals
$\sigma_{\rm pred}$ are conservatively wide relative to the residuals.
KDUQ's curves in panel (b) have close to unit width but are shifted to positive $z$,
including on its own training data.
The Test Corpus in red shows some degradation, with an offset in the median indicating that our model is consistently 
underestimating cross sections relative to data. KDUQ suffers from the same style of degradation, with 
a greater median offset. Recall that this corpus is particularly 
challenging and we do show improvement over the existing standard in total coverage.
Note that the calibration curves here are intended to be comparable with 
Figure 11 of Pruitt \emph{et al.}; our calculation shows better performance of KDUQ here than their own 
paper does. We attribute this difference to the filtering of the KDUQ and Test corpora we perform to 
remove datasets strongly influenced by compound elastic effects.

\mysec{Conclusions}
We have combined a flexible optical potential with a differentiable
scattering solver to quantify uncertainties on the potential and predicted observables. Within a
local Gaussian approximation, we are able to evaluate these uncertainties without
sampling the full parameter posterior. Our method distinguishes between regions 
where the potential is constrained by data and regions where the potential 
is only held by the prior.

The Gaussian-process model improves agreement with held-out cross
sections while achieving nearly nominal coverage with 30\%
narrower predictive intervals than KDUQ. On the separate Test Corpus,
the improvement in fit is smaller and the uncertainties are broader.
Coverage improves over KDUQ but remains below its nominal value,
showing that uncertainty propagation alone does not resolve the
difficulty of predicting beyond the measured region. The deliverable 
in these regions is more trustworthy error bounds rather than a more accurate
mean prediction.

The uncertainty calculation is not specific to optical potentials
or to how the potential is parameterized. It requires the sensitivity of predicted
observables to the model parameters and a local Gaussian approximation
to the posterior. A differentiable solver makes these sensitivities
accessible even when the model involves a numerical solution of the governing equation. This allows
the same approach to quantify uncertainty in other inverse problems
where the physical field of interest cannot be measured directly.

\begin{acknowledgments}
  \emph{Acknowledgments}---This document is approved for unlimited release as LA-UR-26-28459. This work was performed under the auspice of the U.S. Department of Energy by Los Alamos National 
  Laboratory under Contract 89233218CNA000001.  Research reported in this publication was supported 
  by the U.S. Department of Energy LDRD program (Project No. 20250048DR - Assured Artificial Intelligence: 
  Efficient and Scalable Guardrail Certification for Science and Security) at Los Alamos National Laboratory.
  ROS acknowledges support from NSF PHY-2012057, PHY-2309172, and AST-2206321.
  This research used the LANL AI Portal and resources provided by the Information Technology (IT) 
  Division at Los Alamos National Laboratory (Supported by the U.S. Department of Energy, National 
  Nuclear Security Administration under Contract No. 89233218CNA000001).

  KD wrote the scattering solver, the data selection, and the likelihood. KD used Codex
  (OpenAI; GPT-5 series) to parallelize the scattering code and to set up the launching of
  runs, and validated the parallel results against serial runs. KD used Claude (Anthropic; Opus
  4.8, 5, and 5.5) to write the code for the Gaussian-process prior, the posterior bands, the
  calibration statistics, and the figures. KD checked this code against independent
  calculations (dense matrix inversion, finite differences, and synthetic residuals of known
  distribution). KD also used Claude to discuss ideas and the literature and to prepare an early
  draft of the text, which KD and ROS rewrote.
\end{acknowledgments}

\bibliography{refs}

\newpage
\onecolumngrid
\vspace{1.5em}
\begin{center}
  \textbf{\large End Matter}
\end{center}
\twocolumngrid

\mysec{The predictive variance}
Let $\Delta\bm p$ denote a displacement from the fitted parameters.
Linearizing the predicted log cross sections gives
\begin{equation}
  \begin{gathered}
  \ln\hat y_i(\bm p+\Delta\bm p)
  \simeq \ln\hat y_i(\bm p)
  + \sigma_i\sum_a A_{ia}\Delta p_a,
  \\
  A_{ia}=\frac{1}{\sigma_i}
    \frac{\partial\ln\hat y_i}{\partial p_a}.
  \end{gathered}
  \label{em:weightedjacobian}
\end{equation}
The band rests on three approximations; once they hold, the covariance below is exact.
(i)~The measurement errors on $\ln y_i$ are Gaussian and independent, with the variances
$\sigma_i^2$ of Eq.~\eqref{eq:errormodel}.
(ii)~The prior on the parameters $\bm p$ is Gaussian, with precision $\bm\Lambda$.
(iii)~The linearization of Eq.~\eqref{em:weightedjacobian} holds over the width of the posterior.
Here $\bm p$ is
$\bm b=(\bm b^{(1)},\dots,\bm b^{(N_e)})$, which collects the fitted coefficients of
Eq.~\eqref{eq:model} at all $N_e$ experiments, $N_cMN_e$ numbers in total. Assumption (ii) holds
exactly, with $\bm\Lambda^{-1}=K(X,X)\otimes\bm I_{N_cM}$ for the unit-variance prior, where
$K(X,X)$ is the Mat\'ern correlation matrix between the fitted experiments; and because the
potential is affine in $\bm b$, (iii) linearizes only the scattering solver.
Writing $\tilde r_i=r_i/\sigma_i$ for the residuals at the fitted parameters in units of their
uncertainty, (i)--(iii) make the negative log posterior exactly quadratic in $\Delta\bm p$,
\begin{equation}
  \begin{split}
  -2\ln\pi(\bm p+\Delta\bm p)
  ={}&\bigl\lVert\tilde{\bm r}-\bm A\Delta\bm p\bigr\rVert^2\\
  &+(\bm p+\Delta\bm p)^{\!\top}\bm\Lambda(\bm p+\Delta\bm p)+\text{const},
  \end{split}
  \label{em:nlp}
\end{equation}
whose curvature $\bm A^\top\bm A+\bm\Lambda$ does not depend on $\Delta\bm p$. The posterior is
therefore Gaussian, with covariance
\begin{equation}
  \bm\Sigma =
  (\bm A^\top\bm A+\bm\Lambda)^{-1}.
  \label{em:hessian}
\end{equation}
Here $\bm F=\bm A^\top\bm A$ is the Fisher information of Eq.~\eqref{eq:fisher}, and the only
approximation in Eq.~\eqref{em:hessian} is the linearization (iii), which we test below.

In practice we evaluate $\bm\Sigma$ in the dimension of the data rather than of the
parameters. The Woodbury identity~\cite{woodbury1950,hager1989}
(see also Ref.~\cite{rasmussen2006gpml}, Eq.~A.9) gives
\begin{equation}
  \bm\Sigma=\bm\Lambda^{-1}
  -\bm\Lambda^{-1}\bm A^\top
  \bigl(\bm I_N+\bm A\bm\Lambda^{-1}\bm A^\top\bigr)^{-1}
  \bm A\bm\Lambda^{-1},
  \label{em:woodbury}
\end{equation}
which replaces the inverse of a $P\times P$ matrix, $P=N_cMN_e$, with that of an $N\times N$
matrix over the $N=8298$ measured points. The prior covariance
$\bm\Lambda^{-1}$ is known in closed form, so the only factorization needed is of
$\bm I_N+\bm A\bm\Lambda^{-1}\bm A^\top$. The first term is the prior covariance and the second
is the reduction due to the data. The result is exact, with no truncation of the spectrum.
Equation~\eqref{em:woodbury} is an exact rearrangement of
Eq.~\eqref{em:hessian}, not a further approximation: it applies the identity to
$(\bm\Lambda+\bm A^\top\bm I_N\bm A)^{-1}$ with the known $\bm\Lambda^{-1}$. The $N\times N$ matrix
$\bm I_N+\bm A\bm\Lambda^{-1}\bm A^\top$ is the covariance the prior predicts for the
measurements, each divided by its uncertainty, plus the resulting unit noise.

Predictions at a nucleus and energy outside the fit are the standard
GP conditional~\cite{rasmussen2006gpml}. Write $\bm x=(A,Z,E)$ for a standardized input,
$X=\{\bm x_n\}_{n=1}^{N_e}$ for the fitted experiments and $\bm x_\star$ for the predicted experiment;
$K(X,X)$ for the $N_e\times N_e$ Mat\'ern correlation matrix and $K(\bm x_\star,X)$ for the
correlations between $\bm x_\star$ and $X$. Let $\bm b$ be the coefficient vector defined
above, with posterior covariance $\bm\Sigma$ from
Eq.~\eqref{em:woodbury}, and let $\bm\Sigma_0$ be the prior covariance of the $N_cM$
coefficients at one experiment ($\bm\Sigma_0=\bm I$ for the unit-variance prior of
Eq.~\eqref{eq:model}). Conditioning on the fitted coefficients gives the mean of
Ref.~\cite{rasmussen2006gpml}, Eq.~(2.19),
\begin{equation}
  \bar{\bm b}_\star=\bm W\bm b,
  \qquad
  \bm W=K(\bm x_\star,X)K(X,X)^{-1}\otimes\bm I_{N_cM},
  \label{em:gpmean}
\end{equation}
with $\otimes$ the Kronecker product over the coefficient index, and, because those coefficients
are themselves uncertain, a predictive covariance with two terms,
\begin{equation}
  \begin{split}
  \bm\Sigma_\star={}&\bigl[K(\bm x_\star,\bm x_\star)
    -K(\bm x_\star,X)K(X,X)^{-1}K(X,\bm x_\star)\bigr]\bm\Sigma_0\\
    &+\bm W\bm\Sigma\bm W^\top .
  \end{split}
  \label{em:gpcond}
\end{equation}
The first term is the GP conditional variance of Ref.~\cite{rasmussen2006gpml}, Eq.~(2.19); the
second propagates the uncertainty of the values being conditioned on, taking the place of the
i.i.d.\ noise term in their Eq.~(2.24). Far from every fitted experiment $K(\bm x_\star,X)\to0$
and $\bm\Sigma_\star\to K(\bm x_\star,\bm x_\star)\bm\Sigma_0$, the prior.

\mysec{Validity of the linearized variance}
The accuracy of linear uncertainty propagation depends on the
predicted quantity $q$. The potential
is affine in the coefficients, so its variance follows exactly from
their covariance. Cross sections depend nonlinearly on the potential
through the scattering solver and require a separate check.

Let $\Delta\bm p$ be drawn from the Gaussian posterior approximation
with covariance $\bm\Sigma$. Scaling this displacement by a scalar $\epsilon$, so that $\epsilon=1$ samples the full estimated
posterior width and $\epsilon\to0$ approaches the linear limit, gives
\begin{equation}
\begin{split}
  q(\bm p+\epsilon\Delta\bm p)
  ={}& q(\bm p)+\epsilon\nabla q^\top\Delta\bm p\\
     &+\frac{\epsilon^2}{2}
       \Delta\bm p^\top(\nabla^2q)\Delta\bm p
       +O(\epsilon^3).
\end{split}
\end{equation}
The linear term gives the variance
$\epsilon^2\sigma_{\rm lin}^2(q)$ of Eq.~\eqref{eq:band}.
For centered Gaussian displacements, the covariance between the
linear and quadratic terms vanishes. For smooth $q$ with nonzero
linear variance, the sampled width $\sigma_{\rm samp}(q;\epsilon)$, the
standard deviation of $q(\bm p+\epsilon\Delta\bm p)$ over draws of $\Delta\bm p$, therefore satisfies
\begin{equation}
  \frac{\sigma_{\rm samp}(q;\epsilon)}
       {\epsilon\,\sigma_{\rm lin}(q)}
  =1+O(\epsilon^2).
\end{equation}
The size of the correction depends on the derivatives of $q$.

We check this ratio for $q=\ln(\mathrm{d}\sigma/\mathrm{d}\Omega)$
by sampling the coefficient distribution and solving the scattering
problem for each draw. Its approach to unity as $\epsilon$ decreases
checks the small-displacement limit; its value at $\epsilon=1$
tests propagation over the estimated posterior width. This comparison
tests the linearization of the observable, while retaining the
Gaussian approximation to the parameter posterior.

\begin{figure}
  \includegraphics[width=\columnwidth]{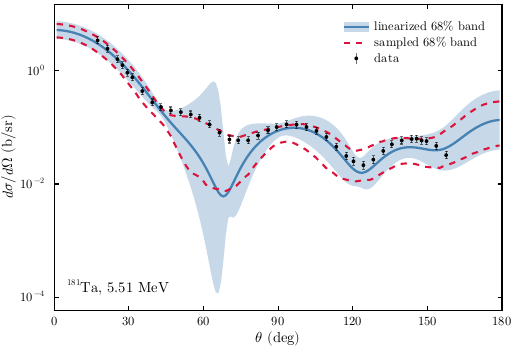}%
  \caption{\label{fig:lin}Linearized and sampled uncertainty
  for neutrons on $^{181}$Ta at 5.51~MeV, a Test Corpus experiment far from the fitted data.
  The filled band is the linearized $\pm1\sigma$ interval, $\ln\hat y\pm\sigma_{\rm lin}$. The
  dashed lines are the 16th and 84th percentiles of the cross section over draws from the
  same posterior, each solved exactly. Error bars on the data combine the reported uncertainty with our measured
  $\dT = 0.137$ (Eq.~\eqref{eq:errormodel}).}
\end{figure}

Figure~\ref{fig:lin} shows an experiment far from the fitted data, where $\sigma_{\rm lin}$ is
large. Near the diffraction minimum the linearized band overstates the spread, and the sampled
distribution is displaced and skewed relative to the fitted curve. The linearization fails in this way once
$\sigma_{\rm lin}$ becomes large, which happens mostly in the Test Corpus. At those points every
calibration result uses the sampled width at $\epsilon=1$ in place of $\sigma_{\rm lin}$.

\mysec{The scaling factor}
The scaling factor $s_c(r)$ sets the size of the allowed deviations
from KD and is fixed before fitting:
\begin{equation}
  s_c(r)=\kappa
  \sqrt{\bigl[\Vkd_c(r)\bigr]^2
    +\bigl[\varepsilon V^{\rm peak}_c f_V(r)\bigr]^2}
  \,T(x).
  \label{eq:scale}
\end{equation}
Here $\kappa$ sets the overall scale, $V^{\rm peak}_c$ is the
peak magnitude of the KD component, and $f_V(r)$ is its volume
Woods--Saxon form factor. The floor factor $\varepsilon$ allows
deviations where the KD component vanishes. The cosine
taper $T(x)$ brings the deviations smoothly to zero over the
outermost part of the radial basis support.

For the unit-variance coefficient prior, the resulting potential
variance at a fixed nucleus and energy is
\begin{equation}
  \operatorname{Var}_0[V_c(r)]
  =s_c(r)^2\sum_{j=1}^{M}\lambda_j\varphi_j(x)^2.
  \label{eq:priorwidth}
\end{equation}
Thus the prior width follows the magnitude of KD where the floor
and taper are negligible, with its radial dependence also set
by the retained GP modes.

\mysec{The dataset-to-dataset scatter $\dT$}
We estimate $\dT$ from independent measurements at nearly identical
nuclei and energies.
We identify 22 pairs of datasets from distinct EXFOR subentries with equal $Z$, $|\Delta A|<1.5$,
$|\Delta E|<0.5$~MeV, and at least five overlapping angles.
We treat the physical differences within these tolerances as
negligible for this estimate.

For a matched pair of datasets $a$ and
$b$, with $\tilde y_b$ denoting dataset $b$ interpolated onto the angles of $a$, we calculate
the mean squared log difference
\begin{equation}
  d_{ab}^2 =
  \frac{1}{|\Theta_{ab}|}
  \sum_{\theta\in\Theta_{ab}}
  \left[\ln y_a(\theta)-\ln\tilde y_b(\theta)\right]^2,
\end{equation}
where $\Theta_{ab}$ contains the angles in the shared range.
Assuming independent errors of equal variance in the two
measurements, the estimated per-dataset scatter is
\begin{equation}
  s_d^2=\frac{1}{2}\langle d_{ab}^2\rangle,
\end{equation}
with the average taken over matched pairs. Removing the
contribution already assigned to reported errors gives
\begin{equation}
  \dT =
  \sqrt{s_d^2-\tilde\sigma_{\rm rep}^2}
  =0.137,
\end{equation}
where $\tilde\sigma_{\rm rep}=0.050$ is the median
reported uncertainty in log cross section over all datasets in the corpus. Using this median
approximates the reported-error contribution to the pair scatter.

We use a single $\dT$ to represent the additional scatter across
datasets and angles, rather than introducing further parameters
for its dependence on nucleus, energy, or angle. This is a
simplification of the error model: the disagreement can include
angular dependence as well as normalization offsets. Unlike a
term fitted alongside the potential, such as that used by KDUQ, this estimate measures
disagreement between experiments rather than residuals of a
particular optical model.

\end{document}